\pdfoutput=1
\documentclass[10pt,conference]{IEEEtran}

\usepackage{booktabs}
\usepackage{multirow}
\usepackage{xcolor}
\usepackage{graphicx}
\usepackage{subcaption}
\usepackage{enumitem}
\usepackage{balance}
\usepackage{cite}
\PassOptionsToPackage{hyphens}{url}
\usepackage[hidelinks]{hyperref}
\usepackage{times}

\usepackage{amssymb}
\newcommand{\system}{\textsc{Mirage}}
\newcommand{\genmodel}{Claude Sonnet}
\newcommand{\opusjudge}{Claude Opus}
\newcommand{\openaijudge}{GPT-5.5}

\makeatletter
\renewcommand\paragraph{\@startsection{paragraph}{4}{2\parindent}%
  {0ex plus 0.1ex minus 0.1ex}{0ex}{\normalfont\normalsize\bfseries}}
\makeatother

\begin{document}

\title{\system{}: Microservice Integration Runtime\\Agent for Generative Emulation}

% TODO: confirm author list / affiliation before posting to arXiv.
\author{\IEEEauthorblockN{XinRan Zhang}\\
\IEEEauthorblockA{University of California, Berkeley\\ zhangxr7@berkeley.edu}}

\maketitle

\begin{abstract}
Integration testing of microservices requires standing in for downstream
dependencies. Existing approaches---record-replay, pattern-mining, and
specification-driven stubs---generate a static artifact before tests run, so
they can only reproduce behaviors encoded at generation time. On the long-tail
scenarios integration testing exists to catch---error paths, boundary
conditions, business logic---these artifacts fail: in our evaluation,
record-replay and pattern-mining reproduce the values a caller asserts on in at
most 1\% of scenarios.

We propose \emph{online LLM simulation}, a paradigm that replaces the
compile-time artifact with request-time reasoning: a language model answers each
dependency request as it arrives, maintaining cross-request state throughout a
test scenario. The model reads the dependency's source code (when available),
caller code, and production traces, then synthesizes behavior on demand---
trading latency ($\sim$1.5--3\,s per request) and cost (\$0.16--\$0.82 per
dependency) for coverage that static artifacts cannot reach.

We instantiate this approach in \system{} and evaluate it on 110 scenarios
across 14 caller--dependency pairs in three microservice systems (Google's
Online Boutique, Weaveworks' Sock Shop, and a custom system), scoring both
structural fidelity and a value-level acceptance metric. With dependency source
(white-box), \system{} reproduces caller-observable values in 93\% of scenarios,
versus at most 1\% for record-replay and pattern-mining and 7\% for opaque
response generation even when it is handed real recorded response bodies; off
the recorded happy path---errors, boundaries, state---the static baselines fall
to near-zero value fidelity in our evaluation, while \system{} remains faithful.
Ablations characterize the approach further: without dependency source
(black-box) \system{} is far weaker than white-box, yet with one example
response body its fidelity improves and stays above the static baselines; the
white-box/black-box gap is driven by \emph{observability} rather than model
capability, so a stronger frontier model helps only marginally. The approach
also holds across dependency source languages and transports, runs on cheaper
open-weight models, is near-deterministic across repeats, and reproduces all
end-to-end caller-test outcomes against the real dependencies.
\end{abstract}

\section{Introduction}
\label{sec:intro}

Integration testing of microservices requires deploying downstream dependencies
with correct configurations, database states, and behavioral semantics. When
dependencies are unavailable or expensive to provision, developers rely on
\emph{dependency simulation}: lightweight surrogates that reproduce the
dependency's behavior during testing~\cite{iyer2024missproblems,atomicjar2023}.
Existing approaches treat this as a \emph{compile-time} artifact-generation
problem: record-replay tools capture request-response pairs from production
traffic~\cite{hoverfly2023}; pattern-mining extracts stateful rules from
traces~\cite{hossain2025stateful}; schema-based tools generate stubs from API
specifications~\cite{prism2023}. All produce a static mock artifact before
tests run.

These approaches share a structural limitation: \textbf{the generated artifact
is coverage-limited by construction.} Any scenario not anticipated during
artifact construction---unseen request parameters, error conditions, cross-request
state dependencies---produces incorrect or missing responses. This matters
because the scenarios most likely to be absent from a trace corpus or a
manually written contract are precisely those that make integration testing
valuable: rare error paths, boundary conditions in business logic, and failure
compensation flows.

As a concrete example, a Sock Shop payment service declines charges above \$100;
on a \$100.01 test charge, record-replay and pattern-mining both return 200
(approved) because they never recorded a decline, whereas the real service
returns \verb|{"authorised": false}|---a boundary condition invisible to
artifact-based approaches (Section~\ref{sec:rq1} gives more).
Such scenarios---error paths, boundary conditions, validation logic---are
routine concerns in integration testing, yet they are structurally
underrepresented in production trace corpora because they occur less frequently
than happy-path interactions. The effect is large: across our three benchmarks,
record-replay and pattern-mining reproduce the values a caller asserts on in at
most 1\% of scenarios, and on error-handling and code-reasoning scenarios
specifically they reproduce almost none.

We propose a different paradigm: \emph{online LLM simulation}.
Rather than compiling a behavioral artifact ahead of time, we keep a language
model in the loop \emph{at test time} and let it synthesize each dependency
response as the request arrives. The model maintains cross-request state
throughout a scenario and reads whatever evidence is available---the
dependency's source code, the caller's code, and production traces---to answer
on demand. No mock specification, recorded corpus, or intermediate
representation is generated in advance. This reframes dependency simulation from
a \emph{retrieval} problem (find the closest recorded behavior) into a
\emph{reasoning} problem (infer the correct behavior), which is what lets it
cover the unseen long tail that defeats static artifacts.

The cost of this shift is an explicit engineering tradeoff: online simulation
takes $\sim$1.5--3\,s per request and \$0.16--\$0.82 per dependency, versus
sub-millisecond static mocks. This makes it a selective tool for CI integration
suites---where behavioral coverage, not per-request latency, is the bottleneck
---rather than a replacement for fast mocks on hot paths.
Figure~\ref{fig:design_space} summarizes the contrast.

We instantiate this approach in \system{}
(\textbf{M}icroservice \textbf{I}ntegration \textbf{R}untime \textbf{A}gent
for \textbf{G}enerative \textbf{E}mulation) and evaluate it on 110 test
scenarios spanning 14 caller-dependency pairs across three microservice systems
of increasing complexity. Our contributions are:

\begin{enumerate}[leftmargin=*]
  \item \textbf{A new paradigm for dependency simulation: online,
    reasoning-based behavioral synthesis.} We move dependency simulation from a
    compile-time artifact (record, mine, or specify ahead of time) to a runtime
    process in which an LLM synthesizes each response on demand while
    maintaining cross-request state, and instantiate it in \system{}.

  \item \textbf{A value-level evaluation showing the paradigm outperforms static
    methods on hard scenarios.} Using a metric that checks the caller-observable
    \emph{values}---not just status codes or response shape---\system{}
    reproduces dependency behavior in 93\% of scenarios with source code and
    18\% without, versus at most 1\% for record-replay and pattern-mining and
    7\% for opaque response generation even when it is given real recorded
    bodies. The gap is widest exactly where integration testing matters: error
    handling and code reasoning, the long-tail behaviors absent from happy-path
    traces, where every static method collapses to near-zero value fidelity.
    Even the source-free black-box mode, while far weaker than white-box,
    exceeds the static baselines.

  \item \textbf{An analysis isolating \emph{observability}, not capability, as
    the white-box/black-box gap.} A signal ablation identifies dependency source
    as the dominant input (97\% alone). We then show that a stronger frontier
    model improves black-box simulation only marginally, whereas supplying one
    example response body raises black-box structural fidelity from 33\% to
    77\%---so the gap is about what the model can observe, not how capable it is.
    Cheaper open models are also competitive, indicating the paradigm is not
    tied to a single frontier model.

  \item \textbf{Operating-envelope characterization.} We report the cost and
    latency tradeoffs that make this a selective testing mode by design
    ($\sim$1.5--3\,s per request, \$0.16--\$0.82 per dependency, $\sim$\$2 per
    14-dependency CI run), show results are near-deterministic across repeats,
    and demonstrate that across 8 end-to-end caller integration tests run
    against the real dependencies and against \system{}, all 8 reach the same
    pass/fail outcome (8/8 agreement).
\end{enumerate}

\begin{figure}[t]
\centering
\scriptsize
\begin{tabular}{@{}p{0.44\columnwidth}p{0.44\columnwidth}@{}}
\toprule
\textbf{Artifact-Based Simulation} & \textbf{Online LLM Simulation} \\
\midrule
Retrieves recorded behavior & Reasons out the correct behavior \\
Static artifact before tests & Response synthesized at request time \\
Coverage-limited on unseen scenarios & Reasons from source code and traces \\
Sub-millisecond latency & $\sim$1.5--3\,s latency per request \\
Low marginal cost & \$0.16--\$0.82 per dependency \\
Deterministic by construction & Near-deterministic in our eval. \\
Best for: anticipated interactions & Best for: long-tail behavioral coverage \\
\bottomrule
\end{tabular}
\caption{Two paradigms for dependency simulation. Online LLM simulation replaces
compile-time retrieval with request-time reasoning, trading latency and cost
for behavioral coverage on the unseen scenarios that defeat static artifacts.}
\label{fig:design_space}
\end{figure}

\noindent
Section~\ref{sec:related} surveys related work.
Section~\ref{sec:approach} defines the approach, benchmarks, and evaluation
protocol. Section~\ref{sec:eval} presents results across four research
questions. Section~\ref{sec:discussion} interprets the findings and discusses
deployment tradeoffs. Section~\ref{sec:threats} addresses threats to validity,
and Section~\ref{sec:conclusion} concludes.

\section{Background and Related Work}
\label{sec:related}

We organize related work along the dependency simulation design space,
from fully static approaches to our proposed runtime synthesis.

\paragraph{Record-Replay and Service Virtualization}
The simplest form of dependency simulation captures HTTP request-response
pairs from live traffic and replays the closest match at test time.
Tools such as HoverFly~\cite{hoverfly2023}, WireMock~\cite{wiremock2023},
and Mountebank~\cite{mountebank2023} implement this approach and are widely
used in industry~\cite{masor2022fake,atomicjar2023,kong2022testing}. They offer
sub-millisecond latency and full determinism \emph{when a test request matches
something in the recorded corpus}, but they have no mechanism to answer a
request they never recorded: on a miss they return a default error, a 404, or an
empty response rather than a plausible one. The very limitation that motivated
research on opaque response \emph{generation}~\cite{du2015interaction,versteeg2016opaque}
is that real integration tests routinely exercise error conditions, boundary
cases, and request parameters absent from the recorded happy-path traffic---
exactly where record-replay ``is limited in accuracy''~\cite{du2015interaction}
and cannot produce a correct response (we quantify this in
Section~\ref{sec:rq1}).

\paragraph{Behavior Mining and Stateful Emulation from Traces}
A richer line of work mines behavior from recorded interaction traces and
synthesizes responses. Hossain et~al.~\cite{hossain2025stateful} mine
contextual dependencies among interaction messages to emulate stateful services,
populating each synthesized response by copying values from the matching
recorded request and from response payloads observed in the trace. Such methods
can reproduce stateful behavior with high fidelity, but under two assumptions
our setting deliberately removes: their recordings are full message captures
that \emph{contain the response bodies} they copy from, and they are evaluated
on traffic drawn from the same distribution as the recording (e.g.,
cross-validation in which every interaction appears in training). When the only
available trace is OpenTelemetry spans, which carry no response bodies, and the
test set is the unseen long tail, a mining baseline has neither bodies to copy
nor in-distribution coverage to exploit. We compare against such a trace-driven
pattern-mining baseline in Section~\ref{sec:rq1} and analyze this gap explicitly.

\paragraph{Schema-Based and Specification-Driven Approaches}
Prism~\cite{prism2023} generates mock servers from OpenAPI specifications,
producing type-correct responses that satisfy interface contracts.
Dredd~\cite{dredd2023} validates API implementations against specifications
rather than generating mocks, but shares the specification-first philosophy.
These tools are \emph{interface-faithful} but not \emph{behavior-faithful}:
they ensure valid response shapes without modeling business logic, state
transitions, or conditional error responses. A specification-driven mock for
a payment service returns 200 for every valid request shape, regardless of
card type or amount---the kind of behavioral reasoning that goes beyond
interface correctness.

\paragraph{Consumer-Driven Contract Testing}
Pact~\cite{pact2023} and Spring Cloud Contract~\cite{springcloudcontract2023}
verify that services honor agreed-upon interaction contracts between consumers
and providers~\cite{robinson2006consumer}. Contract testing detects
interface-level incompatibilities (missing fields, changed types) but does not
synthesize semantically correct behavior for unseen scenarios. A Pact contract
says ``when the consumer sends X, the provider should return Y''---it verifies
\emph{anticipated} interactions rather than simulating \emph{unanticipated}
ones. Our approach is complementary: contract testing checks compatibility of
known interactions; online simulation handles the behavioral long tail.

\paragraph{LLMs for Software Engineering and Testing}
Large language models have been applied to code
generation~\cite{chen2021codex,li2023starcoder,roziere2024codellama},
test generation~\cite{chen2024chatunitest,lemieux2023codamosa}, and program
repair~\cite{xia2023apr}. Most relevant is LLM-based integration test
generation: Pan et~al.~\cite{pan2026saint} combine program analysis with
LLM agents to generate service-level integration tests (SAINT), while
Elumalai~\cite{elumalai2025spring} proposes an LLM workflow for Spring Boot
integration tests. These approaches generate \emph{test code}; \system{}
generates \emph{dependency behavior}---a complementary problem where the LLM
must faithfully simulate a service rather than exercise one.

\paragraph{LLMs as Emulators of Tools and Services}
LLMs have recently been used to stand in for external systems, but with a
different goal than ours. ToolEmu~\cite{ruan2024toolemu} emulates tool execution
to stress-test agents for risky behavior, and DeepAgent~\cite{li2026deepagent}
uses LLM-simulated APIs as a training environment; both drive the emulator from
a natural-language \emph{description} and validate its outputs only for
plausibility, never measuring fidelity against a real service's responses.
\system{} instead grounds synthesis in the dependency's source code or its
traces and is evaluated by value-level fidelity against the real service on
out-of-distribution scenarios. We are not aware of prior work that evaluates an
LLM as a value-level, behavior-faithful dependency simulator for integration
testing; practitioner tools that prompt an LLM to fill OpenAPI mock responses
produce plausibly-shaped data without behavioral grounding or fidelity
evaluation.

Recent work applies LLMs to microservice systems more broadly:
Zhang et~al.~\cite{zhang2025microremed} benchmark LLM capabilities on
microservice remediation, while Yellin~\cite{yellin2025evaluating} finds that
specification complexity degrades LLM performance on microservice
applications---consistent with our design choice to keep an LLM in the loop at
request time rather than commit to a pre-generated behavioral specification.
Wang et~al.~\cite{wang2024llmtest} survey LLM-based software testing methods,
identifying test generation and oracle construction as key directions but
noting that LLM-as-dependency-simulator remains unexplored.
Iyer~\cite{iyer2025aibreaks,iyer2024missproblems} analyzes how AI features
complicate microservice testing and why traditional integration tests miss
real problems, directly motivating the need for richer test doubles like
\system{}.

\paragraph{Distributed Tracing}
Distributed tracing frameworks (OpenTelemetry~\cite{opentelemetry2023},
Jaeger~\cite{jaeger2023}, Zipkin~\cite{zipkin2023}) capture inter-service
communication as spans. Traces have been used for anomaly
detection~\cite{gan2019seer}, performance
diagnosis~\cite{sambasivan2016principled}, and root cause
analysis~\cite{yu2021microrank}. \system{} uses traces as \emph{evidence for
behavioral synthesis}: observed interactions inform the LLM's simulation rather
than serving as monitoring or diagnostic signals.

\paragraph{Behavioral Contracts and Intermediate Representations}
Protocol state machines~\cite{beschastnikh2011leveraging}, session
types~\cite{honda2016session}, and behavioral
contracts~\cite{castagna2009theory} provide formal frameworks for specifying
service behavior. Such representations can, in principle, be generated from
traces and code to yield precise simulators, but they require committing to a
fixed transition model ahead of time. \system{} instead keeps an LLM in the
loop at request time, avoiding a pre-committed behavioral specification.

\paragraph{Positioning}
Table~\ref{tab:positioning} summarizes the positioning. We are not aware of prior
work that combines real-time, per-request LLM dependency simulation with
cross-request state tracking and value-level fidelity measurement. The closest
work~\cite{hossain2025stateful} mines stateful behavior from traces but produces
a static artifact tied to the recorded message contents, so it cannot reason
about unseen scenarios or operate when traces lack response bodies.
\system{} differs by delegating behavioral reasoning to an LLM that operates
at test time with access to source code and traces, reaching higher fidelity in
our evaluation on the non-trivial scenarios where static approaches fall short.

\begin{table}[t]
\caption{Positioning in the dependency simulation design space. \system{}
performs online behavioral synthesis from source code and traces, rather than
producing a pre-generated static artifact.}
\label{tab:positioning}
\centering
\small
\resizebox{\columnwidth}{!}{%
\begin{tabular}{lcccc}
\toprule
\textbf{Approach} & \textbf{Runtime} & \textbf{Source} & \textbf{Stateful} & \textbf{Unseen} \\
\midrule
Record-replay         & Static & ---        & No  & No  \\
Pattern-mining        & Static & ---        & Yes & Limited \\
Schema/contract stubs & Static & Spec       & No  & No  \\
\system{} (ours)      & Online & Code+Trace & Yes & Yes \\
\bottomrule
\end{tabular}}
\end{table}

\section{Approach and Experimental Setup}
\label{sec:approach}

\subsection{Problem and Fidelity Metrics}

The task is simple to state. A caller depends on a downstream service; we want a
simulator that can stand in for that service during the caller's integration
tests. The simulator may use production traces of past interactions and,
when available, the dependency's source code. It is good if the caller cannot
tell it apart from the real service: across a test scenario---a sequence of
calls---every response the simulator returns should be one the real service
could have returned, in both status codes and the values the caller reads.

We measure this with two metrics of increasing strictness. \textbf{Key-level
pass} is structural: a scenario passes if every response is valid JSON, the
whole sequence of HTTP status codes matches the real service, and each response
body matches the real one recursively in keys, value types, and nested shape
(not just top-level keys). This certifies that the caller's deserialization and
control flow behave identically, but does not check the concrete values.
\textbf{Total pass} adds exactly that: the scenario passes at the key level
\emph{and} the caller-observable values are accepted as faithful to the real
response---the right error message, the correct \texttt{authorised} flag, a
converted amount within the service's rounding. Total pass is the headline
metric, since it is what value-level integration assertions depend on; a staged
checker (Section~\ref{sec:protocol}) decides value acceptance. Key-level pass
corresponds to the structural fidelity reported in prior work; total pass is the
value-level criterion this evaluation adds.

\subsection{The \system{} System}

\system{} performs \emph{online LLM-based simulation}: the LLM remains in the
loop at test time, answering each request as it arrives. This is not one-shot
code generation---no standalone simulator artifact is produced. The tradeoff
is higher per-request latency ($\sim$3\,s) and cost (\$0.16--\$0.82 per
dependency) in exchange for behavioral coverage on unseen scenarios---error
conditions, state transitions, and business logic---that no pre-built artifact
anticipated.

\system{} consists of three components:

\paragraph{Context Builder}
Assembles the LLM system prompt from available signals: (1)~dependency source
code (for white-box mode; truncated to 8{,}000 characters),
(2)~caller source code (truncated to 5{,}000 characters), and
(3)~a summary of trace interactions grouped by endpoint and status code
distribution. These character caps are an experimental convenience that keeps
prompts small and cost low for our benchmark services, \emph{not} a design
limit: modern models have context windows measured in the millions of tokens,
so a deployment could supply whole multi-file services without truncation. The
prompt instructs the LLM to simulate the dependency's behavior, track
state across requests, and return responses as JSON with \texttt{status}
(integer), \texttt{body} (object), and optional \texttt{headers} fields.

\paragraph{Per-Request LLM Server}
A FastAPI~\cite{fastapi2023} server that intercepts all HTTP requests to the
mock endpoint. Each incoming request is serialized as a JSON user message
containing the HTTP method, path, query parameters, request body, and relevant
headers (\texttt{Authorization}, \texttt{Idempotency-Key}). The LLM returns a
JSON object with \texttt{status} (integer), \texttt{body} (object or list),
and optional \texttt{headers}. Both messages are appended to a conversation
history that is maintained across requests within a scenario, enabling the LLM
to track accumulated state (e.g., items added to a cart, tokens issued,
reservations pending or confirmed). The conversation is bounded to the last
20 request/response exchanges (40 messages plus the system prompt) to control
context length.
If the LLM returns invalid JSON, one retry with an error correction prompt
is attempted before falling back to a 500 response. A unified client routes
requests across providers; our main results use \genmodel{}, and we report a
cost/capability comparison against cheaper open models (Kimi~K2.5, GLM-5) in
Section~\ref{sec:rq3}.

\paragraph{Scenario Context}
Before each test scenario, the evaluation harness posts the scenario name and
the \emph{sequence of HTTP methods and paths} (but \emph{not} the expected
status codes or response bodies) to a \texttt{/\_\_mock\_\_/scenario} endpoint.
This helps the LLM anticipate multi-step flows (e.g., ``reserve then confirm'')
without leaking the \emph{oracle}---the expected status codes and response
values the scenario is checked against. Between scenarios, a
\texttt{/\_\_mock\_\_/reset} endpoint clears conversation state. We evaluate the
impact of this context in Section~\ref{sec:rq1} via a no-injection ablation.

\subsection{Operating Modes}
\label{sec:modes}

\system{} operates in three signal configurations:

\begin{itemize}[leftmargin=*]
  \item \textbf{White-box}: dependency source code $+$ caller code $+$ traces.
    The LLM has full visibility into $D$'s implementation.
  \item \textbf{Black-box}: caller code $+$ traces (no dependency source).
    Models the common case where the dependency is owned by another team.
  \item \textbf{Ablation variants}: dep-only, caller-only, or traces-only, for
    the signal contribution analysis (Section~\ref{sec:rq3}).
\end{itemize}

\subsection{Baselines}
\label{sec:baselines}

We compare against three static-artifact baselines, spanning the spectrum of how
dependencies are simulated in practice:

\begin{enumerate}[leftmargin=*]
  \item \textbf{Record-replay} (all 14 pairs): captures request-response
    pairs from traces and replays the closest match using path pattern
    normalization and status grouping
    (HoverFly-style~\cite{hoverfly2023}).
  \item \textbf{Pattern-mining} (all 14 pairs): extracts stateful patterns from
    traces, building mock servers with explicit state variables, match
    conditions, and response templates
    (inspired by~\cite{hossain2025stateful}).
  \item \textbf{Opaque response generation} (all 14 pairs): the strongest
    classical \emph{responder}~\cite{du2015interaction,versteeg2016opaque}.
    It is essentially an enhanced form of pattern-mining: it aligns each incoming
    request to its nearest recorded transaction (Needleman--Wunsch sequence
    alignment) and splices request-derived fields into that transaction's
    recorded response, rather than emitting a fixed template. Because it can only
    re-template a \emph{recorded} response, we give it its most favorable
    input---a library of real happy-path response bodies---and also report its
    body-free behavior (Section~\ref{sec:rq1}, Section~\ref{sec:threats}).
\end{enumerate}

\noindent
All three baselines are trace-driven and produce a static artifact before tests
run; they are the direct point of comparison for \system{}'s online synthesis.
We additionally vary the backing LLM (Section~\ref{sec:rq3}) to show the approach
is not tied to one model.

\subsection{Benchmark Systems}
\label{sec:benchmarks}

\begin{table}[t]
\caption{Benchmark systems. OB and SS are community benchmarks originally
written in Go, C\#, Java, and Node.js; we adapt faithful Python replicas that
preserve all endpoint semantics, status codes, and validation logic.}
\label{tab:benchmarks}
\centering
\small
\resizebox{\columnwidth}{!}{%
\begin{tabular}{lrrrrl}
\toprule
\textbf{System} & \textbf{Deps} & \textbf{Pairs} & \textbf{Scenarios} & \textbf{Traces} & \textbf{Origin} \\
\midrule
Demo             & 3 & 3  & 38  & 1{,}234 & Custom \\
Online Boutique  & 6 & 6  & 33  & 1{,}414 & Google \\
Sock Shop        & 5 & 5  & 39  & 1{,}173 & Weaveworks \\
\midrule
\textbf{Total}   & \textbf{14} & \textbf{14} & \textbf{110} & \textbf{3{,}821} & \\
\bottomrule
\end{tabular}}
\end{table}

Table~\ref{tab:benchmarks} summarizes the three benchmark systems. The
\emph{Demo} system is a custom four-service application (order, inventory,
payment, shipping) designed with intentionally tricky patterns: optimistic
locking with version conflicts (inventory), token lifecycle with expiration and
refresh (payment), async polling with status transitions (shipping), and saga
orchestration with compensating actions (order). Its most important property is
that it is original and unpublished, so it appears in \emph{no} model's training
data: white-box fidelity on Demo therefore cannot be explained by memorization
of public code, unlike the two community benchmarks below. We use it as a
controlled, contamination-free stress test where all approaches are compared
under identical conditions.

\emph{Online Boutique}~\cite{onlineboutique2023} is Google's microservice demo
application with 11 services written in Go, C\#, Node.js, and Python. We
target 6 dependency services (product catalog, cart, currency, shipping, email,
payment) called by the checkout and frontend services. Key behaviors include
multi-step checkout (8 chained service calls), credit card validation with
type-specific rejection, and EUR-hub currency conversion with floor rounding.

\emph{Sock Shop}~\cite{sockshop2023} is Weaveworks' microservice benchmark
with 6 services written in Go, Java, and Node.js. We target 5 dependencies
(catalogue, cart, user, payment, shipping) called by the orders and frontend
services. Key behaviors include a \$100 payment threshold, pre-seeded user
accounts with hashed passwords, and tag-based catalogue filtering with
pagination.

For each benchmark, we collect production-style traces via OpenTelemetry
instrumentation~\cite{opentelemetry2023} by running scripted workloads covering
normal flows, error paths, and edge cases. Test scenarios are manually designed
to cover the 9 behavior categories of Table~\ref{tab:categories}
(38 for Demo, 33 for OB, 39 for SS; 110 total), with emphasis on error handling,
code reasoning, and edge cases that are underrepresented in typical production
traces---the unseen scenarios where static mocks fail and where integration
testing provides the most value. The custom Demo system deliberately carries the
heaviest error-handling, code-reasoning, and lifecycle load, since it is the one
benchmark guaranteed free of training-data contamination.

\subsection{Evaluation Protocol}
\label{sec:protocol}

For each test scenario we run the identical call sequence against the real
service and the simulator and compare the captured responses:

\begin{enumerate}[leftmargin=*]
  \item Post scenario context (the call sequence \emph{without} expected
    outcomes) to the simulator's \texttt{/\_\_mock\_\_/scenario} endpoint.
  \item Run the calls against the real service; record its status sequence
    \emph{and the full response bodies}.
  \item Reset the simulator's state via \texttt{/\_\_mock\_\_/reset} and
    re-inject scenario context.
  \item Run the same calls against the simulator; record its statuses and bodies.
  \item Score the scenario with the staged checker below, recording both the
    key-level and total-pass verdicts.
\end{enumerate}

\noindent
Unlike prior dependency-simulation evaluations that retain only status codes or
top-level keys, we persist every real and simulated response body in full, so
value-level fidelity can be scored offline and re-audited.

\paragraph{Staged value checker}
Deciding whether two responses are value-equivalent is delicate: dynamic fields
(timestamps, generated IDs, floating-point rounding) legitimately differ
between runs, so naive equality both under- and over-rejects. We use a
three-stage checker that escalates only when needed:
\begin{enumerate}[leftmargin=*]
  \item \textbf{Hard gate.} Parse, status-sequence match, and recursive
    key/type/shape match. Failing the gate fails the scenario (no value check
    is even attempted). Passing the gate is exactly the \emph{key-level} pass.
  \item \textbf{Conservative pattern oracle.} A set of predicate families
    (numeric equality up to the service's own rounding, set/membership for list
    responses, error-envelope message match, dynamic-field exemption, etc.)
    grounded against the \emph{real} response: a field is checked only if the
    real service exposed it. A byte-identical real/simulated body auto-passes.
    Cases the oracle cannot confidently accept or reject are marked
    \emph{uncertain}.
  \item \textbf{Two-judge LLM panel.} Only \emph{uncertain} cases are escalated
    to a panel of two models from a \emph{different} family than the generator
    (\opusjudge{} and \openaijudge{}), each shown the anonymized real ($A$) and
    simulated ($B$) bodies and a rubric that permits dynamic values to differ.
    Both judges must agree to flip a verdict; both-PASS accepts, both-FAIL
    rejects, and a split is escalated to a human. This keeps the panel
    conservative and prevents same-model self-evaluation.
\end{enumerate}
A scenario counts as a \emph{total pass} if it clears the hard gate and is
accepted by either the pattern oracle or the judge panel. Because the oracle is
grounded on the real response and the judges are independent of the generator,
the checker remains discriminating---it still rejects genuinely wrong values
such as a stale token lifetime or an incorrect converted amount
(Section~\ref{sec:rq1}).

Multi-step scenarios use chained references (e.g., a reservation ID from step~1
is used in step~2's URL). The harness extracts these from response bodies and
substitutes them in subsequent requests for both the real and simulated runs.

\subsection{Implementation Details}

\system{} is implemented in Python using FastAPI for HTTP handling and
LiteLLM for model routing. The Context Builder assembles prompts averaging
3{,}000--5{,}000 tokens (system message), with per-request messages of
200--500 tokens. Conversation history is bounded to 20 exchanges (40 messages
plus system prompt), totaling 6{,}000--10{,}000 tokens per request.

For white-box mode, dependency source files are truncated to 8{,}000 characters
if needed; for multi-file services (original Go/Java source), files are
concatenated with filename headers and truncated to 6{,}000 characters. Caller
source is truncated to 5{,}000 characters. Trace summaries aggregate spans by
endpoint and status code distribution, typically requiring 1{,}000--3{,}000
characters regardless of trace volume.

The evaluation harness supports chained references across calls within a
scenario: a reservation ID returned by call~1 can be substituted into call~2's
URL via template markers (e.g., \texttt{\_\_rsv\_id\_\_}). The harness
performs this substitution identically for both real and simulated services,
ensuring a fair comparison.

All experiments use temperature 0.1 and max\_tokens 2{,}048. Temperature is the
sampling parameter that controls output randomness; we set it low (0.1, near
greedy) so the simulator behaves near-deterministically across repeats. For
reasoning models that do not expose a temperature setting, we omit the parameter
and use the provider default. All conditions (the \system{} signal
configurations of Section~\ref{sec:modes} and the static baselines) use the same
evaluation protocol and metrics: \system{} maintains conversation state across a
scenario and resets between scenarios, pattern-mining uses hand-coded state
variables, and record-replay is stateless.

\paragraph{Data availability}
The code, benchmark adapters, test scenarios, trace corpus, evaluation harness,
LLM prompt templates, and the staged-checker predicate rules are provided as a
replication package.

\paragraph{Trace-test independence}
Test scenarios and production traces are designed independently: traces are
collected from scripted workloads that exercise normal paths (browse, add to
cart, checkout), while test scenarios are manually designed to cover error
handling, edge cases, and code-level reasoning not present in the workload.
For example, the Demo traces contain no double-confirmation or stock-exhaustion
sequences, yet these appear as test scenarios. On OB and SS, the workload
generator runs 5 rounds of the standard user journey; test scenarios include
boundary conditions (e.g., \$100 payment threshold in SS, expired card in OB)
that the workload does not exercise. This independence means the LLM cannot
simply memorize trace patterns to pass the evaluation.

\paragraph{Scenario context as potential leakage}
The scenario context tells the LLM \emph{what calls will arrive} (e.g.,
``POST /items/item-001/reserve then POST /reservations/\{id\}/confirm'') but
\emph{not what status codes to return}. This is analogous to a human developer
reading a test plan before writing a mock. We evaluate the impact of this
context by running a no-injection ablation in Section~\ref{sec:rq1}.

\section{Evaluation}
\label{sec:eval}

We organize the evaluation around four research questions that operationalize
the paper's single thesis---that request-time reasoning reproduces dependency
behavior where compile-time retrieval cannot:

\begin{itemize}[leftmargin=*]
  \item \textbf{RQ1 (Fidelity)}: Does online LLM simulation reproduce
    caller-observable dependency behavior more faithfully than static
    record-replay, pattern-mining, and opaque response generation, and where do
    the approaches diverge?
  \item \textbf{RQ2 (Generality)}: Does the fidelity advantage hold across
    benchmarks, dependency source languages, transport protocols, and repeated
    runs?
  \item \textbf{RQ3 (What drives fidelity)}: Which input signals matter most, is
    the white-box/black-box gap a limit of \emph{model capability} or of
    \emph{observability}, and how much does the backing model matter?
  \item \textbf{RQ4 (Practicality)}: What are the cost and latency implications
    for CI use?
\end{itemize}

\noindent
Throughout, we report the two metrics of Section~\ref{sec:approach}:
\emph{key-level pass} (parseable, full status-sequence match, and recursive
key/type/shape match) and \emph{total pass} (key-level \emph{plus}
caller-observable values accepted by the staged checker). Total pass is the
headline number. \system{} cells are the mean of three repeated runs; the
static baselines are deterministic. No scenario injection is used in any
reported result (we verify injection does not inflate fidelity below).
The 110 scenarios are split unevenly across the 14 pairs (column \textbf{N} in
Table~\ref{tab:main}, from 1 to 19, summing to 110), so each cell is a
percentage over that pair's own N rather than over 110; the per-category
breakdown (Table~\ref{tab:categories}) instead pools all 110 scenarios.

\subsection{RQ1: \system{} vs.\ Static Baselines}
\label{sec:rq1}

\begin{figure}[t]
\centering
\includegraphics[width=0.95\columnwidth]{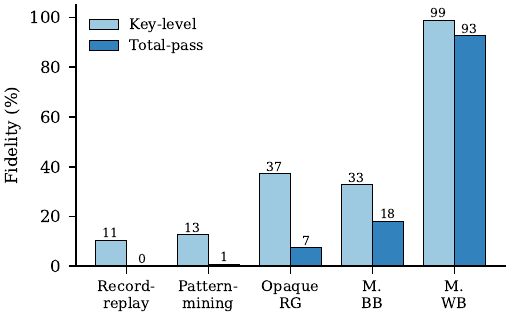}
\caption{Mean fidelity across the 14 pairs (key-level and total-pass). The static
methods reconstruct some response \emph{structure}---opaque RG, given real
bodies, most of all---but almost never reproduce the caller-observable
\emph{values} (total-pass $\le$7\%). \system{} white-box reaches 93\% total-pass;
even its source-free black-box mode (18\%) exceeds every static method.}
\label{fig:main}
\end{figure}

\begin{table}[t]
\caption{Main result: fidelity across all 14 caller--dependency pairs
(110 scenarios). \textbf{N} is the number of test scenarios for that pair;
each method cell is key-level\,/\,total-pass, \emph{expressed as a percentage of
that pair's own N} (so for $N{=}1$ a cell is either 0 or 100, and 67 means
$2/3$). \textbf{Opaque} is opaque response generation given its most favorable
input (a recorded library of real happy-path response bodies); \system{} runs
are the mean of three repeats. The \textbf{Mean} row is the unweighted mean
across the 14 pairs (each pair counts equally regardless of its N).
\textbf{WB} = white-box (dependency source $+$ caller $+$ traces);
\textbf{BB} = black-box (caller $+$ traces, no dependency source).}
\label{tab:main}
\centering
\small
\setlength{\tabcolsep}{4pt}
\resizebox{\columnwidth}{!}{%
\begin{tabular}{llcccccc}
\toprule
 & & & \multicolumn{5}{c}{\textbf{key-level\,/\,total-pass}} \\
\cmidrule(lr){4-8}
\textbf{Bench} & \textbf{Dependency} & \textbf{N} & \textbf{Replay} & \textbf{Pattern} & \textbf{Opaque} & \textbf{\system{} BB} & \textbf{\system{} WB} \\
\midrule
\multirow{3}{*}{Demo}
 & inventory  & 19 & 32/0\% & 37/5\% & 11/11\% & 21/16\% & 100/95\% \\
 & payment    & 13 & 0/0\%  & 8/8\%  & 0/0\%   & 0/0\%   & 85/69\% \\
 & shipping   & 6  & 17/0\% & 33/0\% & 0/0\%   & 33/0\%  & 100/100\% \\
\midrule
\multirow{6}{*}{OB}
 & product-cat. & 7 & 0/0\% & 0/0\% & 43/0\%  & 29/14\%  & 100/100\% \\
 & cart         & 6 & 0/0\% & 0/0\% & 100/17\%& 100/100\%& 100/100\% \\
 & currency     & 7 & 0/0\% & 0/0\% & 57/0\%  & 57/14\%  & 100/76\% \\
 & shipping     & 4 & 0/0\% & 0/0\% & 50/25\% & 25/0\%   & 100/100\% \\
 & email        & 1 & 100/0\%& 100/0\%& 100/0\%& 0/0\%   & 100/100\% \\
 & payment      & 8 & 0/0\% & 0/0\% & 38/0\%  & 38/38\%  & 100/100\% \\
\midrule
\multirow{5}{*}{SS}
 & catalogue    & 10 & 0/0\% & 0/0\% & 0/0\%   & 33/0\%   & 100/90\% \\
 & cart         & 7  & 0/0\% & 0/0\% & 14/14\% & 14/14\%  & 100/100\% \\
 & user         & 11 & 0/0\% & 0/0\% & 9/0\%   & 9/9\%    & 100/100\% \\
 & payment      & 8  & 0/0\% & 0/0\% & 100/38\%& 88/50\%  & 100/100\% \\
 & shipping     & 3  & 0/0\% & 0/0\% & 0/0\%   & 11/0\%   & 100/67\% \\
\midrule
\textbf{Mean} & (14 pairs) & 110 & \textbf{11/0\%} & \textbf{13/1\%} & \textbf{37/7\%} & \textbf{33/18\%} & \textbf{99/93\%} \\
\bottomrule
\end{tabular}}
\end{table}

Table~\ref{tab:main} presents the main result. Averaged over the 14 pairs,
\system{} white-box achieves \textbf{99\% key-level and 93\% total-pass}
fidelity, versus \textbf{0--1\% total-pass} for record-replay and
pattern-mining and \textbf{7\% total-pass} for opaque response generation even
when the latter is handed real recorded response bodies. The pattern beneath
these numbers is the paper's central observation. Each static baseline is
bounded by what it \emph{recorded or anticipated} ahead of time: it returns a
faithful response only when the request resembles one already in its corpus or
contract. When a test exercises behavior outside that envelope---an unrecorded
error, an unseen boundary, a state-dependent conflict---its value-level fidelity
falls toward zero. \system{} replaces this \emph{retrieval} of recorded behavior
with request-time \emph{reasoning} over the dependency's code and traces, and in
our evaluation is the only approach that stays faithful off the recorded happy
path. It wins or ties on value-level total-pass on all 14 pairs.

\paragraph{Why the static baselines score low}
Two checks confirm the low scores reflect a genuine ceiling rather than weak
implementations. First, the OpenTelemetry traces for Online Boutique and Sock
Shop record inter-service \emph{spans} (method, path, status) but \emph{not}
response bodies---the standard situation for production tracing---so a
trace-driven artifact has no recorded body to replay or mine and must serve a
placeholder that fails the recursive value check by construction. Second, and
more tellingly, we implemented opaque response
generation~\cite{du2015interaction,versteeg2016opaque}---the strongest classical
\emph{responder}, which aligns each incoming request to its nearest recorded
transaction and splices request-derived fields into that recorded
response---and gave it its most favorable input: a library of \emph{real}
happy-path response bodies (the \textbf{Opaque} column). With real bodies to
template, it recovers structure better than pattern-mining (mean key-level 37\%
vs.\ 13\%); yet its value-level total-pass is only 7\%, and---as the category
breakdown shows next---\emph{every} one of those passes falls on the happy-path
operations it had recorded. A method that can only re-emit a recorded response
cannot answer a request whose correct response was never recorded, no matter how
well it matches. (Section~\ref{sec:threats} reports the body-free variant, which
necessarily degrades to record-replay.)

\paragraph{Black-box is weaker, but still exceeds the static baselines}
Without dependency source, \system{} (BB) drops to 33\%\,/\,18\%---an order of
magnitude below white-box, but still well above replay and pattern-mining. Its
strength is uneven and informative: it reaches 100\% on OB cart (a CRUD service
whose behavior is fully implied by the caller's usage) but only 9\% on SS user
(which depends on pre-seeded account data the black box cannot observe). We
dissect this gap in RQ3.

\paragraph{Category-level breakdown}
Table~\ref{tab:categories} pools all 110 scenarios by behavior category, and the
same pattern holds within each. The static methods concentrate their
value-level passes on categories with recorded or templatable behavior---a
little on \texttt{basic\_crud} and \texttt{generalization} (opaque RG, with real
bodies, reaches 14\% and 33\%), with only isolated passes in
\texttt{error\_handling} and \texttt{code\_reasoning}. On the categories that lie
off the recorded happy path---\texttt{stateful\_lifecycle}, \texttt{pagination},
\texttt{auth\_lifecycle}, \texttt{async\_lifecycle}---all three static
approaches score \textbf{0\% total-pass}, opaque RG's real-body advantage
included; even on \texttt{error\_handling} they manage at most 11\% (a single
pattern-matched 401), against \system{}'s 95\%. These are precisely the
behaviors integration testing exists to exercise, and they are unanticipated by
construction. \system{} (WB) holds at 75--100\% across the same categories,
because it infers the correct response (a 409 conflict, a declined charge,
a validation 400) rather than retrieving one it has seen.

\begin{table}[t]
\caption{Fidelity by behavior category, all 110 scenarios pooled.
\textbf{N} is the number of scenarios in that category (summing to 110); each
method cell is key-level\,/\,total-pass as a percentage of that category's N.
\textbf{Opaque} = opaque response generation with real happy-path bodies. On the
stateful, pagination, auth-, and async-lifecycle categories, all three static
methods reach 0\% total-pass; \system{} retains nonzero value fidelity in every
category.}
\label{tab:categories}
\centering
\small
\setlength{\tabcolsep}{4pt}
\resizebox{\columnwidth}{!}{%
\begin{tabular}{lcccccc}
\toprule
 & & \multicolumn{5}{c}{\textbf{key-level\,/\,total-pass}} \\
\cmidrule(lr){3-7}
\textbf{Category} & \textbf{N} & \textbf{Replay} & \textbf{Pattern} & \textbf{Opaque} & \textbf{\system{} BB} & \textbf{\system{} WB} \\
\midrule
basic\_crud         & 27 & 3/0\%   & 3/0\%   & 38/14\% & 43/24\% & 100/97\% \\
error\_handling     & 19 & 0/0\%   & 11/11\% & 11/0\%  & 32/11\% & 100/95\% \\
code\_reasoning     & 19 & 5/0\%   & 5/0\%   & 42/11\% & 39/21\% & 95/84\% \\
stateful\_lifecycle & 18 & 17/0\%  & 22/0\%  & 22/0\%  & 22/22\% & 100/100\% \\
auth\_lifecycle     & 4  & 0/0\%   & 0/0\%   & 0/0\%   & 0/0\%   & 100/75\% \\
async\_lifecycle    & 2  & 50/0\%  & 50/0\%  & 0/0\%   & 0/0\%   & 100/100\% \\
pagination          & 3  & 0/0\%   & 0/0\%   & 0/0\%   & 0/0\%   & 100/67\% \\
generalization      & 6  & 17/0\%  & 17/0\%  & 67/33\% & 50/33\% & 100/89\% \\
stress\_tests       & 12 & 10/0\%  & 10/0\%  & 20/0\%  & 20/10\% & 90/90\% \\
\bottomrule
\end{tabular}}
\end{table}

\paragraph{Qualitative examples}
Three cases illustrate the pattern. In OB's bad-card checkout, an Amex number
(starting with~3) should produce a 400 (\texttt{UnacceptedCreditCard}); replay
instead returns a recorded Visa 200, while \system{} (WB) reads the validation
logic. In SS's payment-boundary test, a \$100.01 charge should return
\texttt{authorised=false}, matching the \$100 threshold. In Demo's
version-conflict test, an outdated reservation version should return a 409 from
the optimistic-locking check. Each is a value- or status-level behavior that no
trace-derived artifact reproduces.

\paragraph{The checker is discriminating, not lenient}
Total-pass is gated by the staged checker, which rejects genuinely wrong values
even from \system{}. Of \system{} (WB)'s 330 scenario-runs, 324 pass at the key
level but only 304 pass totally: the 20 residual key-level passes that fail the
value check are real infidelities the oracle correctly catches---e.g.\ a stale
payment-token lifetime (\texttt{expires\_in}) or a currency conversion off by a
sub-cent rounding step. Value acceptance is attributed to the conservative
pattern oracle (277 cases) or the two-judge panel (24), with 3 human-adjudicated;
the panel both confirms faithful dynamic-value differences and rejects wrong
ones, so the 93\% figure reflects screened, not merely structural, agreement.

\paragraph{Scenario context does not inflate results}
All reported numbers use \emph{no} scenario injection. Re-running three
configurations with and without the call-sequence hint changes nothing
materially (e.g.\ Demo inventory 19/19 without vs.\ 18/19 with injection---the
single regression is injection \emph{mis}leading the model), confirming the
fidelity comes from reasoning, not from leaked oracle information.

\smallskip\noindent\textbf{Answer to RQ1.}
Across 14 pairs, \system{} (WB) reaches 99\%/93\% key-level/total-pass, versus
0--1\% total-pass for record-replay and pattern-mining and 7\% for opaque
response generation given real recorded bodies. The static baselines score their
passes on recorded happy-path categories and 0\% total-pass on the
off-happy-path behaviors integration testing targets; \system{}---which reasons
rather than retrieves---is the only approach in our evaluation that stays
faithful there, and even its source-free black-box mode (33\%/18\%) exceeds the
static baselines.

\subsection{RQ2: Generality Across Benchmarks, Languages, Transports, and Runs}
\label{sec:rq2}

\paragraph{Across benchmarks}
White-box fidelity is stable across the three systems despite their differing
complexity, origin languages (Go, C\#, Java, Node.js), and behavioral patterns
(optimistic locking, async polling, saga orchestration): per-benchmark white-box
total-pass is 87\% (Demo), 95\% (OB), and 95\% (SS). Black-box, in contrast,
varies with how much each service's behavior is externally observable
(Demo 8\%, OB 33\%, SS 15\% total-pass)---again pointing to observability, not
model behavior, as the source of variance (RQ3).

\paragraph{Across source languages}
Because our benchmark services are Python, one might worry that white-box
fidelity is inflated by Python being unusually easy for an LLM to read. We test
this directly. We hand-translated the demo inventory service (optimistic
locking, version conflicts, cursor pagination, structured error envelopes) into
behavior-equivalent \textbf{Go} and \textbf{Java}, and fed each as the
white-box source while leaving the running service---and therefore the real
responses and the oracle---unchanged, isolating the effect of source language.
Fidelity is identical across all three: 100\%\,/\,95\% key-level/total-pass for
Python, Go, and Java alike, on every repeat. White-box fidelity comes from
reasoning about the service's \emph{logic}---what triggers a 409, how the
version counter advances---which is expressed equivalently across languages, not
from any property of Python.

\paragraph{Across transports}
A complementary worry is that our results are tied to HTTP/JSON. We therefore
re-ran the inventory dependency over two \emph{real}, non-HTTP transports: gRPC
(genuine protobuf over an HTTP/2 channel, with gRPC status codes mapped to their
HTTP equivalents on both the real service and the mock) and a message queue
(genuine ZeroMQ request--reply, where the caller publishes a request and blocks
for a reply). Only the wire changes; the dependency logic, scenarios, and
value-level oracle are held fixed. White-box fidelity is essentially unchanged
across all three transports---100\%\,/\,95\% (HTTP), 95\%\,/\,89\% (gRPC), and
95\%\,/\,89\% (message queue) key-level/total-pass---and the white-box advantage
over black-box is preserved on every transport. As with source language,
\system{} reproduces the dependency's behavior from its \emph{logic}, not from
the protocol carrying it. (The small gRPC/MQ shortfall is two genuine
value-level misses the judges correctly rejected, not a transport artifact.)

\paragraph{Across repeated runs}
At temperature~0.1 the approach is near-deterministic: across the 14 pairs and
three repeats, key-level and total-pass rates are identical in 26 of 28 cells,
with the remaining two varying by a single scenario. Failures are systematic
(missing information), not stochastic noise, so run-to-run variation in the
main table is small.

\smallskip\noindent\textbf{Answer to RQ2.}
The fidelity advantage is general. White-box total-pass is consistent across
the three benchmarks (87--95\%), identical across Python/Go/Java source, and
$\approx$95\% across HTTP, gRPC, and message-queue transports, with the
white-box advantage over black-box preserved throughout; results are
near-deterministic across repeats (26/28 cells identical). In this study, the
service logic matters more than the language or protocol that carries it.

\subsection{RQ3: What Drives Fidelity---Observability or Capability?}
\label{sec:rq3}

\paragraph{Signal ablation}
Table~\ref{tab:ablation} isolates each input signal (Sonnet, all 110 scenarios
pooled; the full-WB row reads 98\%\,/\,92\% here versus the 99\%\,/\,93\%
headline because the latter is the per-pair mean over three repeats, while the
ablation pools all scenarios from a single run).
Dependency source code is by far the largest signal: alone it yields
97\%\,/\,75\%, nearly matching full white-box (98\%\,/\,92\%). Caller code alone
(35\%\,/\,16\%) or trace status-histograms alone (18\%\,/\,7\%) are far weaker,
and black-box---caller $+$ traces, the realistic no-source setting---lands at
33\%\,/\,18\% (Table~\ref{tab:ablation}). The ordering is intuitive: the
response a dependency returns is determined by its own implementation, so seeing
that implementation is worth more than any amount of external observation.

\begin{table}[t]
\caption{Signal ablation (Sonnet, 110 scenarios, key-level\,/\,total-pass \%).
Dependency source is the dominant signal; without it, caller code and traces
recover only a fraction of fidelity, but adding one example response body to the
black box (last row) closes most of the gap.}
\label{tab:ablation}
\centering
\small
\resizebox{\columnwidth}{!}{%
\begin{tabular}{llcc}
\toprule
\textbf{Condition} & \textbf{Signals} & \textbf{Key} & \textbf{Total} \\
\midrule
\system{} (WB)   & dep $+$ caller $+$ traces & 98 & 92 \\
dep-only         & dep source only           & 97 & 75 \\
\system{} (BB)   & caller $+$ traces         & 33 & 18 \\
caller-only      & caller source only        & 35 & 16 \\
traces-only      & status-histogram traces   & 18 & 7  \\
\midrule
\system{} (BB) $+$ ex.\ body & caller $+$ traces $+$ 1 example body & 77 & 40 \\
\bottomrule
\end{tabular}}
\end{table}

\paragraph{Is the black-box gap a model limit or an information limit?}
We next test whether a \emph{stronger} model would close the black-box gap.
First, we re-ran \emph{every} black-box scenario that Sonnet failed (90
scenarios) with two stronger, independent frontier models (\opusjudge{} and
\openaijudge{}, each three repeats); together they fixed only 11---and 79 were
fixed by neither, so model scale is not the limiting factor here. Second, we
gave the \emph{same} Sonnet black box one piece of additional
\emph{observability}: a single redacted example response body per endpoint
(structure only, dynamic values stripped, no request$\rightarrow$response
mapping; harvested leak-safely from recorded real responses). This
``richer-trace'' condition lifts black-box key-level fidelity from 33\% to
\textbf{77\%} (last row of Table~\ref{tab:ablation}). Of the 79 failures the
stronger models could not fix, this one example body recovers \textbf{71\%}
(56/79) at the key level---behaviors model scale alone did not produce from the
black-box signals.

Both probes point the same way: \textbf{in this benchmark the black-box gap is
driven more by observability than by model scale.} A bigger model barely helps a
black box that cannot see the wire format, whereas one structural example
recovers most of the lost fidelity---so the cheapest way to improve black-box is
to surface response structure (e.g.\ via body-capturing traces) rather than reach
for a larger model. The white-box failures tell the mirror-image story: stronger
models fix 6 of Sonnet's 9, leaving 3 payment-harness artifacts, so white-box is
already near its ceiling and its residual \emph{is} largely a capability limit.

\paragraph{How much does the backing model matter?}
We also ran the full matrix with two cheaper open models (key-level\,/\,total-pass,
3 repeats). All three are competitive on white-box \emph{structure}, but the
cheaper models lose value-level fidelity and degrade more in black-box, where
less of the answer is spelled out by the source: white-box\,/\,black-box is
98/92\%\,/\,33/18\% for Claude Sonnet, 91/72\%\,/\,18/6\% for Kimi~K2.5, and
95/73\%\,/\,21/7\% for GLM-5. The picture is bounded on both ends---cheaper open
models still reach 72--73\% white-box total-pass, while the stronger
\emph{frontier} models of the probe above improve black-box only marginally---so
the achievable range across the tested model spectrum is narrow relative to the
effect of \emph{information} (white-box vs.\ black-box, or adding an example
body). The main results use Sonnet, which is both strong and, per RQ4,
inexpensive.

\smallskip\noindent\textbf{Answer to RQ3.}
Dependency source is the dominant signal (dep-only 97\%/75\%, near full
white-box). For the black-box failures in this benchmark, observability matters
more than model scale: of 90 black-box failures two stronger frontier models fix
only 11, yet one example response body recovers 71\% of the 79 they could not
fix. Across the cheap-to-frontier model spectrum the achievable range is narrow
next to the effect of information---what the model can observe matters more than
which model it is.

\subsection{RQ4: Cost, Latency, and Deployment}
\label{sec:rq4}

\system{} trades latency and cost for behavioral coverage; it is a selective
testing mode, not a faster mock. Each Sonnet call takes $\sim$1.5\,s
(Opus $\sim$3\,s) and, with $\sim$1.7 calls per scenario, the full 110-scenario
suite runs in a few minutes. The entire evaluation---the main matrix in
triplicate plus the judge panel and all ablations---cost on the order of
\textbf{\$25} at list prices, dominated by the white-box runs; a single CI pass
over 14 dependencies is roughly \$2 with Sonnet.

Per-request latency is $\sim$1.5\,s (Sonnet) to $\sim$3\,s (Opus), or
$\sim$2.5--5\,s per scenario ($\sim$1.7 calls) and $\sim$7.5--15\,s cumulative at
serial depth~5; cost per dependency is \$0.16 (Sonnet) to \$0.82 (Opus) for the
full suite, i.e.\ \$2.24--\$11.48 per 14-dependency CI run, and each mock process
uses $\sim$200\,MB versus $\sim$2\,GB for the real service. For shallow workflows
(serial depth $\le$3) latency stays within typical CI timeouts; for deep chains,
mitigations include using Sonnet or applying \system{} selectively to complex
dependencies while keeping fast static mocks for simple ones. The approach needs
no dependency containers, databases, or staging infrastructure, so its net CI
footprint depends on the pipeline it replaces (Section~\ref{sec:discussion}
develops this tradeoff).

\paragraph{End-to-end test equivalence}
To check that \system{} preserves the \emph{outcomes} a caller's integration
tests depend on, we ran the Demo OrderService caller against both the real
dependencies and \system{} mocks across 8 end-to-end scenarios (happy paths,
declined-card compensation, insufficient-stock propagation, zero-quantity edge
cases, multi-order stateful flows). White-box \system{} produces \textbf{8/8}
matching caller outcomes---the caller's tests pass or fail identically whether
backed by real services or \system{}.

\smallskip\noindent\textbf{Answer to RQ4.}
Cost is \$0.16--\$0.82 per dependency (\$2.24 per CI run with Sonnet) at
$\sim$1.5--3\,s per call---practical for CI integration suites at serial
depth $\le$3, not for latency-sensitive unit tests. White-box \system{}
reproduces 8/8 end-to-end caller outcomes against the real dependencies.

\section{Discussion}
\label{sec:discussion}

\paragraph{Artifact coverage and trace representativeness}
Our results show that artifact-based approaches achieve low fidelity on
error-handling and code-reasoning scenarios under the trace volumes in our
study. This is not an artifact of weak baselines---even opaque response
generation, given real recorded bodies, collapses to 0\% value fidelity on these
categories---but a structural consequence of \emph{retrieving} recorded behavior:
a response that was never recorded cannot be replayed, mined, or spliced, only
\emph{reasoned} about. That said, these scenarios are underrepresented in
typical production traces, and we have not tested whether larger body-carrying
trace corpora would narrow the gap; advanced mining on denser, body-carrying
trace samples could improve artifact-based coverage for some scenario classes.

\paragraph{Deployment guidance}
We suggest a few heuristics, offered as guidelines from a three-benchmark study
rather than universal thresholds. Use white-box (99\%/93\%) when dependency
source is available and error/edge-case coverage matters; use black-box
(33\%/18\%) as a fallback for control-flow assertions when source is unavailable,
keeping in mind it is reliable where the caller's usage or an example response
pins down the wire format and unreliable where the response depends on data the
dependency holds privately (e.g.\ pre-seeded accounts); and keep static mocks for
anticipated, latency-critical interactions. A hybrid split---static mocks for
simple dependencies, \system{} for behaviorally complex ones---captures most of
the fidelity gain at lower cost (e.g., \$0.48 with Sonnet for 3 of 10
dependencies).

\paragraph{If dependency source alone gets 97\%, why use traces?}
Dependency source is the dominant signal (dep-only 97\%/75\%), but our small
benchmarks understate the value of traces: they enable the black-box setting
when source is unavailable, act as a compressed behavioral proxy when source
exceeds the context window, and capture runtime configuration not evident from
code. Source alone suffices on services that fit in context; the value of traces
grows with service size and with the realism of the no-source setting.

\paragraph{Value-level fidelity and remaining error}
Unlike evaluations that stop at structural agreement, our total-pass metric
checks the values the caller observes, via a real-grounded pattern oracle and an
independent two-judge panel (Section~\ref{sec:approach}). The residual white-box
error is small and concrete: of 330 white-box runs, 324 pass structurally and
304 totally, with the 20-run gap being genuine value bugs the checker
catches---a stale token lifetime, a sub-cent currency-rounding mismatch. These
are simulation errors, not metric artifacts, and they bound how far value-level
fidelity can be pushed on these services. Most benchmark scenarios communicate
over HTTP/JSON; our transport probe (Section~\ref{sec:rq2}) covers one inventory
dependency over gRPC and ZeroMQ request--reply, but broader gRPC, message-queue,
and event-driven systems remain untested.

\paragraph{Onboarding effort}
Setting up \system{} for a new benchmark took roughly 8 hours across our three
systems---collecting traces, writing scenarios, and configuring the runner,
dominated by scenario design---with no manual prompt writing, as the Context
Builder assembles prompts automatically from source and trace summaries.

\section{Threats to Validity}
\label{sec:threats}

\paragraph{Internal validity}
Scenario injection could inflate fidelity, so we report all results
\emph{without} it; the no-injection ablation (Section~\ref{sec:rq1}) shows
fidelity is equal or higher without it. The judge panel could in principle be
lenient, but it is grounded on the real response, uses two judges from a
different model family than the generator, and requires unanimity; it still
rejects 20 of \system{}'s structurally-correct white-box runs as genuine value
bugs, so it screens values rather than passing on structure alone. The LLM may
have seen Online Boutique and Sock Shop during pre-training, but the original,
unpublished Demo system shows the same white-box$\rightarrow$black-box pattern,
indicating results are not driven by memorization.

\paragraph{External validity}
All three benchmarks use HTTP/JSON and are adapted as Python services from
polyglot originals. The concern that reading Python inflates white-box fidelity
is addressed by the source-language ablation (Section~\ref{sec:rq2}):
behavior-equivalent Go and Java source yield identical fidelity, and black-box
mode---our conservative number---uses no dependency source at all. The study
covers three systems and 110 scenarios but is not exhaustive; enterprise systems
may have more dependencies and richer interaction patterns. Baseline strength is
itself a threat, addressed by running all baselines on all 14 pairs and adding
opaque response generation~\cite{du2015interaction,versteeg2016opaque} under its
most favorable input (Section~\ref{sec:rq1}); the low baseline numbers reflect a
property of the problem---unanticipated behavior cannot be retrieved from a
recording that never contained it---not weak implementations.

\paragraph{Construct validity}
Fidelity is defined relative to our test scenario set, which is not exhaustive
(e.g., no rate limiting, circuit breaking, or distributed transactions), and
category labels are assigned manually. Our total-pass metric treats a value as
faithful if it is one the real service could return; it does not capture
non-functional properties such as latency or headers the caller does not read.

\section{Conclusion}
\label{sec:conclusion}

We presented an empirical study of online LLM simulation for microservice
dependency testing: instead of generating a static artifact before tests run,
the LLM answers each dependency request at test time from source code, caller
context, and production traces while maintaining cross-request state. Measured by
whether the caller-observable \emph{values} match the real service, \system{}
reproduces dependency behavior in 93\% of 110 scenarios with source code and
18\% without, versus at most 1\% for record-replay and pattern-mining and 7\%
for opaque response generation. The key distinction is where behavior is
produced: static methods \emph{retrieve} recorded or anticipated behavior, so off
the recorded happy path---errors, boundaries, state---their value fidelity
collapses, whereas request-time \emph{reasoning} remains faithful, exactly on the
error-handling and code-reasoning scenarios integration testing exists to catch.

Two findings shape deployment. Dependency source is the dominant signal, so
white-box is preferred when source is available; and the white-box/black-box gap
is one of \emph{observability}, not model capability---stronger models barely
help a source-free mock, but a single example response body more than doubles its
structural fidelity, a low-cost lever for the common no-source case. At
$\sim$1.5--3\,s per request and \$0.16--\$0.82 per dependency, the approach
complements rather than replaces fast static mocks on well-understood
interactions. Future work will surface response structure to black-box
simulation automatically, broaden the transport study, and scale to enterprise
systems with many dependencies.

\balance
\bibliographystyle{IEEEtran}
\bibliography{references}

\end{document}